\documentclass{aa}
\usepackage{graphicx}
\usepackage{txfonts}
\usepackage{natbib}
\usepackage[hidelinks]{hyperref}
\usepackage{xcolor}
\hypersetup{
    colorlinks,
    linkcolor={red!50!black},
    citecolor={blue!50!black},
    urlcolor={blue!80!black}
}

\begin{document}

\title{Super lithium-rich K giant with low $^{12}\rm{C}$ to $^{13}\rm{C}$ ratio
\thanks{Based on observations obtained with the Apache Point Observatory 3.5-metre telescope, which is owned and operated by the Astrophysical Research Consortium.}}

\author{Y.T. Zhou \inst{1,2}
   \and J.R. Shi \inst{1,2}
   \and H.L. Yan \inst{1}
   \and Q. Gao \inst{1}
   \and J.B. Zhang \inst{1,2}
   \and G. Zhao \inst{1,2}
   \and K.Pan \inst{1,3}
   \and Y. B. Kumar \inst{1}
   }

\institute{Key Laboratory of Optical Astronomy, National Astronomical Observatories, Chinese Academy of Sciences, Beijing 100012, China. \\
\email{sjr@nao.cas.cn}
\and University of Chinese Academy of Sciences, Beijing 100049, China.
\and Apache Point Observatory and New Mexico State University, P.O. Box 59, Sunspot, NM, 88349-0059, USA.}
\date{Received 03 January 2017/ Accepted 8 March 2018}

\abstract
   {The lithium abundances in a few percent of giants exceed the value predicted by the standard stellar evolution models, and the mechanisms of Li enhancement are still under debate. The Large Sky Area Multi-Object Fiber Spectroscopic Telescope (LAMOST) survey has obtained over six million spectra in the past five years, and thus provides a great opportunity to search these rare objects and to more clearly understand the mechanisms of Li enhancement.}
   {The aim of this work is to accurately measure the Li abundance and investigate the possible mechanisms of Li enrichment for a newly found super Li-rich giant, TYC\,3251-581-1, located near the luminosity function bump with a low carbon isotopic ratio.}
   {Based on the high-resolution spectrum we obtained the stellar parameters ($T_\mathrm{eff}$, $\log g$, [Fe/H]), and determined the elemental abundances of Li, C, N, $\alpha$, Fe-peak, r-process, s-process elements, and the projected rotational velocity. For a better understanding of the effect of mixing processes, we also derived the $^{12}\rm{C}$ to $^{13}\rm{C}$ ratio, and constrained the evolutionary status of TYC\,3251-581-1 based on the BaSTI stellar isochrones.}
   {The super Li-rich giant TYC\,3251-581-1 has $\rm{A(Li)} = 3.51$, the average abundance of two lithium lines at $\lambda = 6708$ \AA\ and 6104 \AA\ based on the non-local thermodynamic equilibrium (NLTE) analysis. The atmospheric parameters show that our target locates on the luminosity function bump. The low carbon isotopic ratio ($^{12}\rm{C}/^{13}\rm{C} = 9.0 $), a slow rotational velocity $v\sin i = 2.2\ \rm{km\,s}^{-1}$, and no sign of IR excess suggest that additional mixing after first dredge up (FDU) should occur to bring internal synthesized Li to the surface. The low carbon ($[\rm{C}/\rm{Fe}] \sim -0.34$ ) and enhanced nitrogen ($[\rm{N}/\rm{Fe}] \sim 0.33$) are also consistent with the sign of mixing.}
   {Given the evolutionary stage of TYC\,3251-581-1 with the relatively low $^{12}\rm{C}/^{13}\rm{C}$, the internal production which replenishes Li in the outer layer is the most likely origin of Li enhancement for this star.}


\maketitle

\section{Introducion}

It is predicted that lithium is destroyed considerably throughout the lifetime for the giants, the dilution of lithium occurs when the convective envelope circulates the surface materials to the deep regions where the temperature is high enough (about several million Kelvin) to destroy Li. According to the standard stellar models, the significant depletion of surface Li takes place during the first dredge-up (FDU) in low mass stars \citep[e.g.][]{Iben67a, Iben67b}. The standard models predicted that the Li abundance of low-mass stars with solar metallicity is less than about 1.5 dex when the star evolves to the end of the FDU on the red giant branch (RGB, \citealp{Charbonnel10}), and it is suggested that the Li abundance would be further declined after the RGB bump because of the burning by the extra mixing \citep[e.g.][]{Charbonnel07, Charbonnel10}. However, numerous observations show that a few percent of giants hold an Li abundance higher than 1.5 dex \citep[e.g.][]{Brown89, dela97, Kumar11, Adamow14}. These so-called Li-rich stars provide a challenge to the standard stellar evolution models.

After the discovery of the first Li-rich K giant star \citep{Wallerstein82}, various scenarios, including the effects of external sources and internal production, have been envisaged to interpret such `anomalous' abundance. Suggested external sources include contamination by the ejecta of nearby novae \citep{Martin94}, mass transfer from the Li-enhanced companion \citep{Kirby16}, and the engulfment of planets and/or brown dwarfs \citep{Alexander1967, Siess99}. On the other hand, \citet{Sackmann99} proposed that Li is freshly synthesized inside stars known as cool bottom processing, in this process, $^{3}\rm{He}$ enriched after the FDU can be burned into $^{7}\rm{Be}$ at the regions that have high enough temperature, $^{7}\rm{Be}$ is then transported to a cooler zone, where it captures an electron and decays into $^{7}\rm{Li}$ \citep[][hereafter, CF mechanism]{Cameron71}.

Several studies have been devoted to investigating the issue of reconciling the existence of Li-rich giants. \citet{Charbonnel00} proposed that Li-rich giants could be formed at the specific evolutionary stages, that is, the RGB bump for low mass stars, and the early asymptotic giant branch (AGB) for intermediate-mass stars. In such cases, when the hydrogen burning shell in the stars erases the discontinuity in the mean molecular weight left behind at the end of FDU, the extra-mixing is expected to trigger the CF mechanism to produce fresh Li. Nevertheless, \citet{Kumar11} found the Li-rich giants clustered around the red clump and the horizontal branch, then they suggested that the onset of CF mechanism might locate at these evolutionary stages. However, other observational results indicate that Li-rich stars can be found almost anywhere along the RGB \citep{Monaco11, Lebzelter12, Martell13}. Among $\sim170$ of the Li-rich giants included the Li-rich collection of \citet{Casey16} with seven new-found by \citet{Li18}, only $\sim20$ super Li-rich giants \citep{Adamow15} with Li abundance higher than 3.3 dex \footnote[1]{the Li abundance in meteorite \citep{Asplund09}} have been found up to now \citep[e.g.][]{Kumar09, Martell13, Monaco14, Casey16}.

In the past five years LAMOST has obtained spectra for six million stars, and the number is still increasing \citep{Cui12, Zhao12}. It is an ideal database in which to seek Li-rich giant candidates because of the huge number of spectra \citep{Silva14, Li16}. In this paper, we report a new discovery of super Li-rich giant from the LAMOST survey, TYC\,3251-581-1, and its follow-up observations with the 3.5 m telescope at the Apache Point Observatory (APO). The paper is organized as follows. In Sect. 2 we briefly describe the observation and data reduction, and in Sect. 3 we present the details of the determinations of the stellar parameters and elemental abundances. Finally, we discuss the potential scenarios of Li enhancement based on the evolutionary status of this star in Sect. 4.

\section{Observations}
This candidate Li-rich giant was found with the LAMOST low resolution spectrum, and selected based on the visual inspection of the spectrum and the measurement of equivalent width for the Li resonance line. The high-resolution and high S/N spectra for TYC\,3251-581-1 were obtained by the cross-dispersed \'{e}chelle spectrograph mounted on the 3.5 m telescope at the APO. The wavelength coverage is from 3200 to 10000 \AA\ with a resolution of $\sim$31,500. The star was observed three times with a total integration time of 1.5 hours on Oct. 22, 2014. The S/N is $\sim160$ at 6708 \AA.

The observed data were reduced with the revised automatic IDL programme, which was designed for the FOCES \'{e}chelle spectrograph \citep{pfe98}. The standard procedures were applied to extract one-dimensional spectra, and the uncertainty in wavelength calibration is less than 0.01 \AA.

\section{Spectroscopic analysis}
\subsection{Atmospheric parameters}
The stellar parameters were determined with a standard procedure based on the MAFAGS-OS atmospheric model \citep{gru09}. The effective temperature ($4670\pm80$ K) was estimated by requesting the same abundance from the neutral iron lines with different excitation potentials, which is in good agreement with the photometric temperature of 4650 K (V-K) obtained by the calibrations of \citet{Alonso99}.
The spectroscopic surface gravity is derived by minimizing the difference between the abundances measured from \ion{Fe}{i} and \ion{Fe}{ii} lines. The result is consistent with the trigonometric gravity ($\log g_{Gaia}$), which is based on the parallax of $\sim 0.58\pm0.30$ mas reported by the recently released Gaia DR1 \citep{Gaia16}. The trigonometric gravity is calculated through
\begin{equation}
\mathrm{log}\ g = \mathrm{log}\ g_{\odot} + \mathrm{log}(\frac{M}{M_{\odot}}) + 4*\mathrm{log}(\frac{T_\mathrm{eff}}{{T_\mathrm{eff}}_{\odot}}) + 0.4*(M_{bol} - {M_{bol}}_{\odot})
,\end{equation}
where the absolute bolometric magnitude is estimated by the formula $M_{bol} = V_{mag} + 5*\mathrm{log}\ \pi + 5.0 - A_{V} + BC$, the interstellar extinction is denoted as $A_{V}$, and the bolometric correction is derived by the empirical relations of \citet{Alonso99}. The microturbulent velocity was determined by requiring the abundance from \ion{Fe}{i} lines being independent of their reduced equivalent widths. The derived stellar parameters are given in Table \ref{tab1}, along with the estimated errors in the stellar parameters. About 50 well separated and mild ($20 \sim 100$ m\AA\ ) \ion{Fe}{i} lines and eight \ion{Fe}{ii} lines were used in the parameter determination.

Stellar mass and age were derived based on the Gaia parallax, spectroscopic $T_\mathrm{eff}$ combined with the BaSTI Ver.5.0.1 evolutionary tracks \citep{Pietrinferni04}. We employed the absolute bolometric magnitude $M_\mathrm{bol}$ and $T_\mathrm{eff}$ to match the theoretical calculations in the evolutionary tracks, which is based on the scaled solar metal mixture. In the process, the canonical model without overshooting was adopted. Following \citet{Pietrinferni04}, the theoretical tracks with exact metallicity Z = 0.014 were computed by interpolating on a point-by-point basis among the available grid of metallicities. The uncertainties were estimated by taking into account the errors of $M_\mathrm{bol}$ and $T_\mathrm{eff}$. The stellar radius was derived by $R=\sqrt{GM_\mathrm{\star}/g}$.

Following \citet{Adibekyan12}, we calculated the spatial velocity components, $(\rm{U}_\mathrm{LSR}, \rm{V}_\mathrm{LSR}, \rm{W}_\mathrm{LSR})$ = (34, -23, -16) $\rm{km\,s^{-1}}$, and the distance away from the galactic plane $Z = 0.43$ kpc. According to the criteria of \citet{Reddy06}, TYC\,3251-581-1 likely belongs to the thin disc with a probability of 98\%.

\begin{table}
\tabcolsep10pt
\scriptsize
\begin{center}
\caption[1]{Stellar parameters, kinematics information for TYC 3251-581-1 \label{tab1}}
\begin{tabular}{lll}
\hline\hline\noalign{\smallskip}
 Property &  Value & Reference\\
\noalign{\smallskip}
\hline\noalign{\smallskip}
 Position (J2000)                         & R.A. \quad \quad   00:20:36.35  & \\
                                          & DEC. \quad        +46:59:05.05  & \\
 Magnitudes \quad  B                      & 11.94                           & \citet{Hog00}\\
 \hspace{1.4cm}  V                        & 11.04                           & \citet{Hog00}\\
 \hspace{1.4cm} $\rm{K}_\mathrm{2MASS}$   & 8.256                           & \citet{Cutri13}\\
 $\pi$ [mas]                              & $ 0.58\pm0.30$                  & \citet{Gaia16}\\
 $\rm{V}_\mathrm{helio}$                  & $-34.68\pm0.06$                 & This work \\
 $\rm{U}_\mathrm{LSR} \ [km\,s^{-1}]$     & $ 34.3 $                        & This work\\
 $\rm{V}_\mathrm{LSR}$                    & $-23.51$                        & This work\\
 $\rm{W}_\mathrm{LSR}$                    & $-15.82$                        & This work\\
 \noalign{\smallskip} \hline
 $\rm{[Fe/H]}$                            & $-0.09\pm0.10$                 & This work\\
 $T_\mathrm{eff}$ \ [K]                   & $ 4670\pm80  $                 & This work\\
 $\log g \ \rm{[cm\,s^{-2}]} $            & $ 2.30\pm0.20$                 & This work\\
 $\log g_{Gaia}\ \rm{[cm\,s^{-2}]} $      & $ 2.28\pm0.50$                 & This work\\
 $\xi_\mathrm{t} \ [\mathrm{km\,s^{-1}}]$ & $ 1.57\pm0.20$                 & This work\\
 $v\sin i  \  \rm{[km\,s^{-1}]} $         & $ 2.2 \pm0.3 $                 & This work\\
 Mass $[M_\odot]$                         & $2.16 \pm1.85$                 & This work\\
 Age [Gyr]                                & $0.72 \pm5.30$                 & This work\\
 Log $(L/L_\odot)$                        & $2.12 \pm0.50$                 & This work\\
 Radius $[R_\odot]$                       & $17.27\pm8.40$                 & This work\\
\noalign{\smallskip} \hline
\end{tabular}
\end{center}
\end{table}

\begin{figure}
 \centering
   \includegraphics[width=8.0cm]{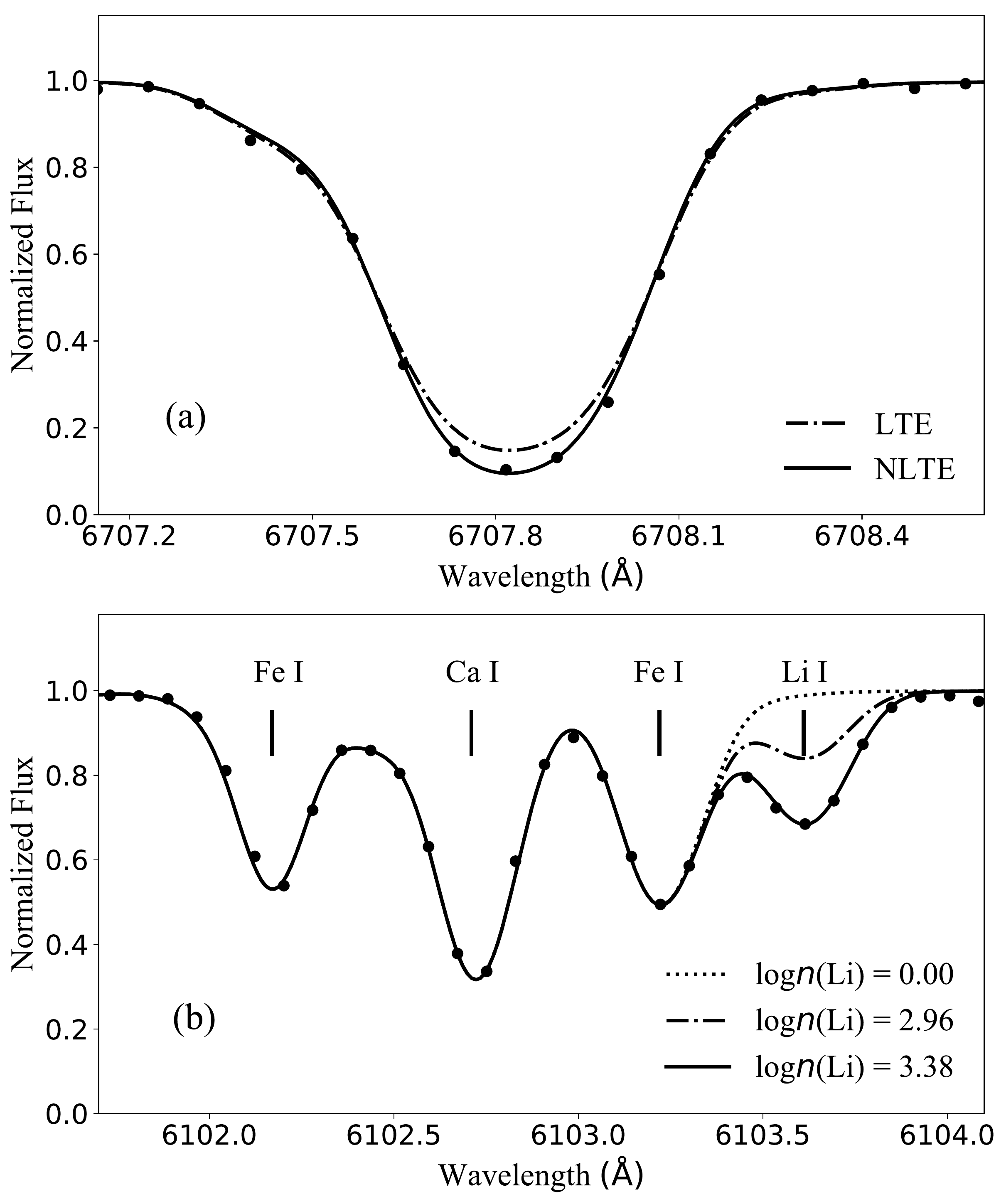}
      \caption{Spectral synthesis of \ion{Li}{i} line at 6708 \AA\ (panel a), and the subordinate line at 6104 \AA\ (panel b) for TYC 3251-581-1.}
         \label{figure1}
   \end{figure}

\subsection{Determinations of abundances}

The Li abundance analysis was performed using the spectral synthesis method with the SIU software package developed by \citet{Reetz91} for the Li resonance line at $\lambda = 6708$ \AA\ and the subordinate line at 6104 \AA. The contributions from the other lines nearby have been considered, and the lithium abundance was obtained until the best fit was reached between the synthetic and observed spectra.
For the resonance line, a local thermodynamic equilibrium (LTE) Li abundance is 3.68 dex by employing the line list compiled by \citet{Carlberg12}. Though the subordinate line at 6104 \AA\ is usually very weak for the normal giants, it is strong in our target, an LTE abundance of $\rm{A(Li)}=3.38$ dex (where $\rm{A(Li)}=\log[n(\rm{Li})/n(\rm{H})]+12$) is derived by adopting a line list from VALD database \citep{Kupka99}. The uncertainties in the derived lithium abundances that arose because of the errors of the atmospheric parameters were evaluated to be 0.26 dex and 0.13 dex for Li lines at $\lambda = 6708$ \AA\ and $\lambda = 6104$ \AA\ (Table \ref{tab2}). We note that the synthetic spectrum at 6708 \AA\ cannot reach the core of this line, as shown by the dashed line in Fig. \ref{figure1}. The high lithium giants suffer a large non-local thermodynamic equilibrium (NLTE) effect \citep{Lind09}, we investigated the NLTE effects for the two lines based on the Li atomic model of \citet{shi07}. When the NLTE effects included, our results show that the NLTE corrections\footnote[2]{The NLTE correction is denoted as: $\Delta_\mathrm{NLTE} = \log n\mathrm{(Li)}_\mathrm{NLTE} - \log n\mathrm{(Li)}_\mathrm{LTE}$} from the two lines at $\lambda = 6708$ \AA\ and $\lambda = 6104$ \AA\ are -0.20, +0.15 dex, respectively. The NLTE lithium abundances of 3.48 and 3.53 are obtained, leading to a smaller difference between two lines compared with those from the LTE analysis. The average of the NLTE abundances from the two lines was adopted as the final Li abundance with A(Li) = 3.51 dex. We also interpolated the NLTE corrections by using the grid of \citet{Lind09}, the corrections for the two lines at $\lambda = 6708$ \AA\ and $\lambda = 6104$ \AA\ are -0.22 and +0.18 dex, respectively. Their results are in good agreement with ours. In the process of NLTE analysis, we measured the uncertainties of NLTE correction due to errors of atmospheric parameters based on the method of \citet{shi07} (see Table \ref{table: 3}). The differential NLTE correction is slightly larger of the resonance line than that of the subordinate line caused by the changes of stellar parameters, and the errors in $T_\mathrm{eff}$ and $\log g$ have a major effect on the uncertainties of NLTE correction.

The carbon abundances were derived from the synthesis of the \ion{C}{i} line at 5380 \AA, and the C$_\mathrm{2}$ Swan bandheads at 5135 \AA\ \citep{Alexeeva15}. Since the neutral nitrogen lines are either too weak or blended by the telluric absorption lines, we fixed the carbon abundance and determined the nitrogen abundance based on the CN bands near 8003 \AA. The [C/Fe] $= -0.34$ and [N/Fe] $=0.33$ show that our target is deficient in carbon, and enhanced in nitrogen (Table \ref{tab2}).

\begin{figure}
    \centering
    \includegraphics[width=9.0cm]{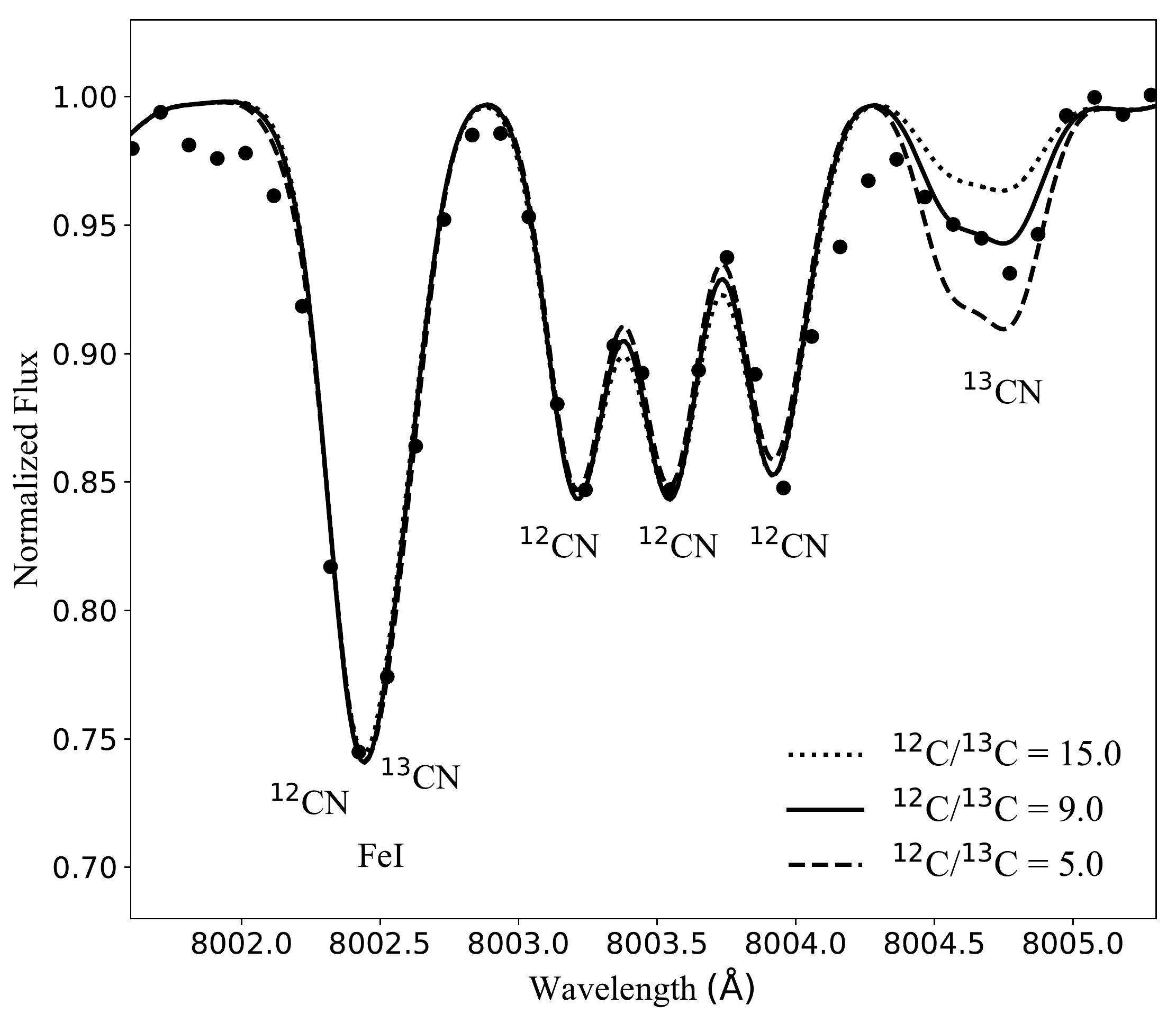}
      \caption{Determination of the carbon isotopic ratio $^{12}\rm{C}$ to $^{13}\rm{C}$ for TYC\,3251-581-1, The three synthetic spectra show ratios of 15 (dotted), 9 (solid, best fit) and 5 (dashed), respectively.}
         \label{figure2}
\end{figure}

The interpretation of the source being Li-rich is likely to be associated with the extra mixing. The mixing process carries the internal material, which has undergone the CNO cycle to the surface, leading to a decrease in $^{12}\rm{C}$ and increase in $^{13}\rm{C}$. It is suggested that the carbon isotopic ratio ($^{12}\rm{C} $ to$ ^{13}\rm{C}$) is a good indicator for tracing the stellar mixing process \citep{Eggleton08, Charbonnel10}. We obtained the $^{12}\rm{C}/^{13}\rm{C}$ value by spectral synthesis for a small group of molecular lines from the region near 8003 \AA, as shown in Fig. \ref{figure2}. The line list used for spectral synthesis is from \citet{Carlberg12} and the derived $^{12}\rm{C}/^{13}\rm{C}$ is $\sim$ 9.0, which is much lower than that of the typical $^{12}\rm{C}/^{13}\rm{C}$  ($\sim25$) after the FDU \citep[e.g.][]{Eggleton08, Palmerini11}.

For purpose of comparison (see Sec. 4), we also measured the abundances of the odd-Z (Na, Al), $\alpha$ (Mg, Si, Ca, and Ti), Fe-peak (Sc, V, Cr, Mn, Co, Ni, and Zn), and neutron capture elements (Y, Ba, La, Nd, Eu) with the spectral synthesis method, and the detailed chemical abundances are reported in Table \ref{tab2}. The derived abundances of $\alpha$ elements ([Mg/Fe], [Si/Fe], and [Ti/Fe]) are 0.06 dex, 0.03 dex, and -0.09 dex, respectively, which are in agreement with that of the typical thin disc star \citep{Adibekyan11}. This is consistent with the result from the kinematical information. The line list for the determination of chemical abundances is compiled from the \citet{Zhao16}, \citet{Jacobson13} and \citet{Neves09}.

The projected rotation velocity is helpful to distinguish the episodes from different sources, such as the internal and external origin. Following \citet{Carlberg12}, we derived the projected rotational velocity of TYC\,3251-581-1 by fitting the unblended \ion{Fe}{i} lines at 6750.2, 6733.2, and 6726.7 \AA. We derived a low projected rotational velocity of 2.2 $\rm{km\,s}^{-1}$ using this method.

\section{Discussion}

\begin{figure}
  \centering
  \includegraphics[width=9.0cm]{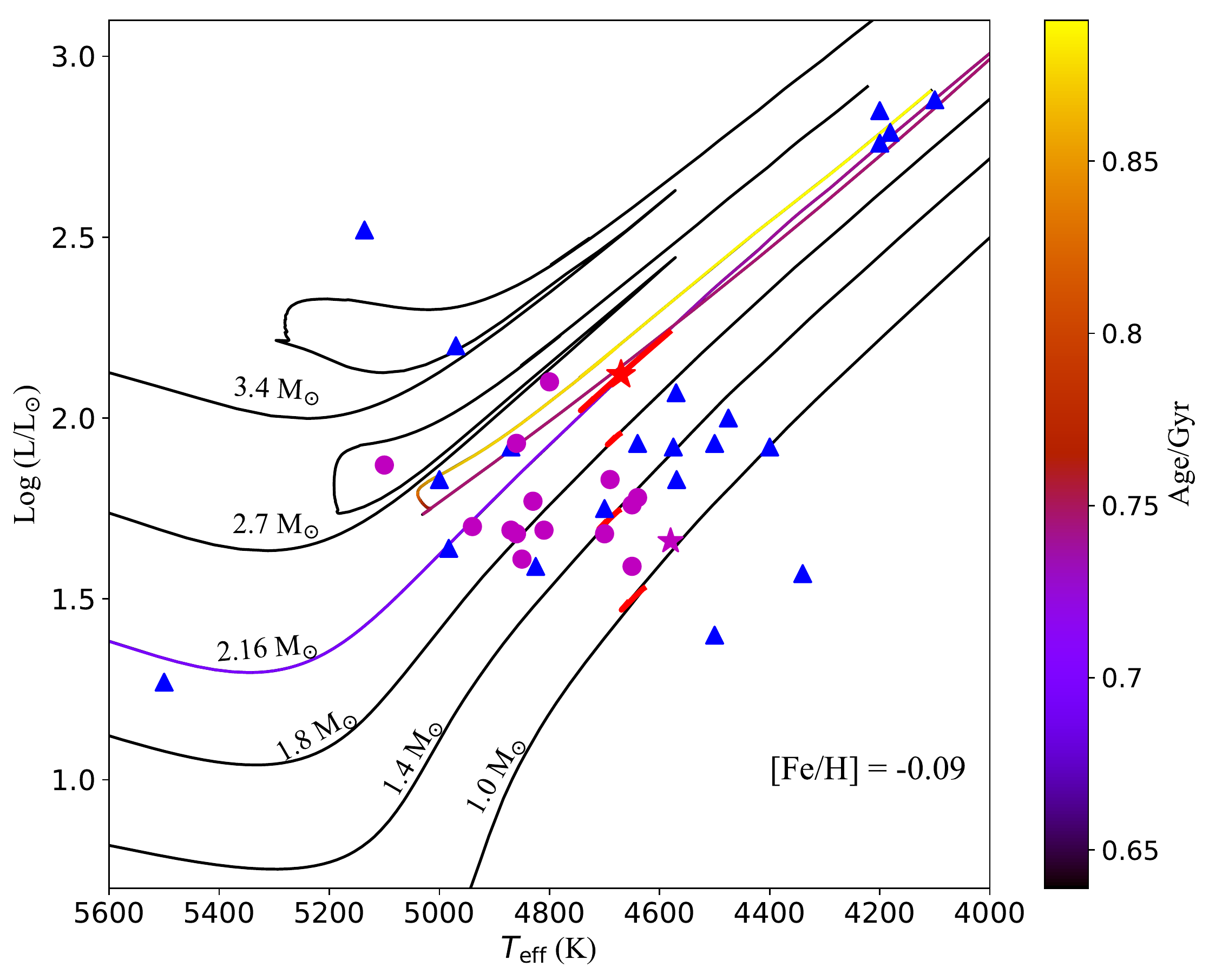}
     \caption{TYC 3251-581-1 is indicated as a red star in the BaSTI tracks with metallicity Z = 0.014 \citep{Pietrinferni04}, and the magenta star is the star HD 77361 \citep{Kumar09}. The \citet{Charbonnel00} sample with [Fe/H] = -0.46 $\sim$ 0.10 dex is denoted as the blue triangles, while the magenta points show the \citet{Kumar11} sample with metallicities ranging from -0.27 to 0.18 dex. The location of the RGB bump is indicated by the red bold line at each track.}
     \label{figure3}
\end{figure}

According to the BaSTI tracks \citep{Pietrinferni04}, TYC\,3251-581-1 is close to the region of RGB bump in Hertzsprung-Russell diagram (see Fig. \ref{figure3}) with a mass M = 2.16 $\mathrm{M_\odot}$ and age = 0.72 Gyr. Regarding the Li rich giants at the RGB bump, \citet{Charbonnel00} suggested that the enhanced Li caused by the extra mixing would again be decreased when the carbon isotopic ratio drops due to the deeper mixing processes. It implies that the enriched Li would take place prior the signature of reduction of $^{12}\rm{C}/^{13}\rm{C}$. This scenario may coincide with the Li-rich giants with high $^{12}\rm{C}/^{13}\rm{C}$ ($\ge\ \sim20$), but it is not consistent with our derived super Li-rich giant with a low $^{12}\rm{C}/^{13}\rm{C}$ of $\sim9.0$. Recently \citet{Kumar11} discovered a dozen Li-rich giant with low $^{12}\rm{C}/^{13}\rm{C}$, and the star HD\,77361 is also located at the RGB bump.

One of the external mechanisms to increase the Li abundance of a star is to engulf a planet or a brown dwarf, as first suggested by \citet{Alexander1967}, which probably happens anywhere along the RGB. To reach the meteoritic Li abundance, an accreted planet of two Jupiter masses (M$_\mathrm{Jup}$) should have an Li abundance of $\rm{A(Li)} \sim 6$, based on the assumptions of \citet{Carlberg10}, this planet is so anomalous that Li abundance is much higher than the meteoritic abundance of the solar system.
Moreover, \citet{Siess99} modelled the accretion of substellar companions (SSCs) by the solar-mass RGB star. They explored the possibly observational features for engulfing SSCs included IR emission, Li overabundance, and rotational velocity. All observed signatures are compatible with the accretion of SSCs, however the difficulty their scenario encountered is that the very high Li abundance requires a huge amount of Li accretion, so it can not properly account for the very high Li abundance ($> 2.8$ dex).
Recently, \citet{Aguilera16a} considered the different typical masses and various metallicities of the giants for this episode, their calculations showed that the SSCs with mass higher than 15 M$_\mathrm{Jup}$ would be dissolved in the radiative zone rather than in the convective envelope. Thus, if there is no any extra mixing, the largest Li abundance could be achieved at 2.2 dex by engulfing the SSCs with a mass of 15 M$_\mathrm{Jup}$. In the case of external pollution, the external sources increase not only the Li materials, other light element abundances ($^{6}\mathrm{Li}$, $^{9}\mathrm{Be}$ and $^{11}\mathrm{B}$) and isotope ratio ($^{12}\rm{C}/^{13}\rm{C}$) \citep{Siess99, Israelian01} but also the orbital angular momentum. This scenario is unlikely for TYC\,3251-581-1 due to it being super Li-rich, the low carbon isotopic ratio, and $v\sin i \sim 2.2 \, \rm{km\,s}^{-1}$.

 \begin{figure}
   \centering
   \includegraphics[width=9.0cm]{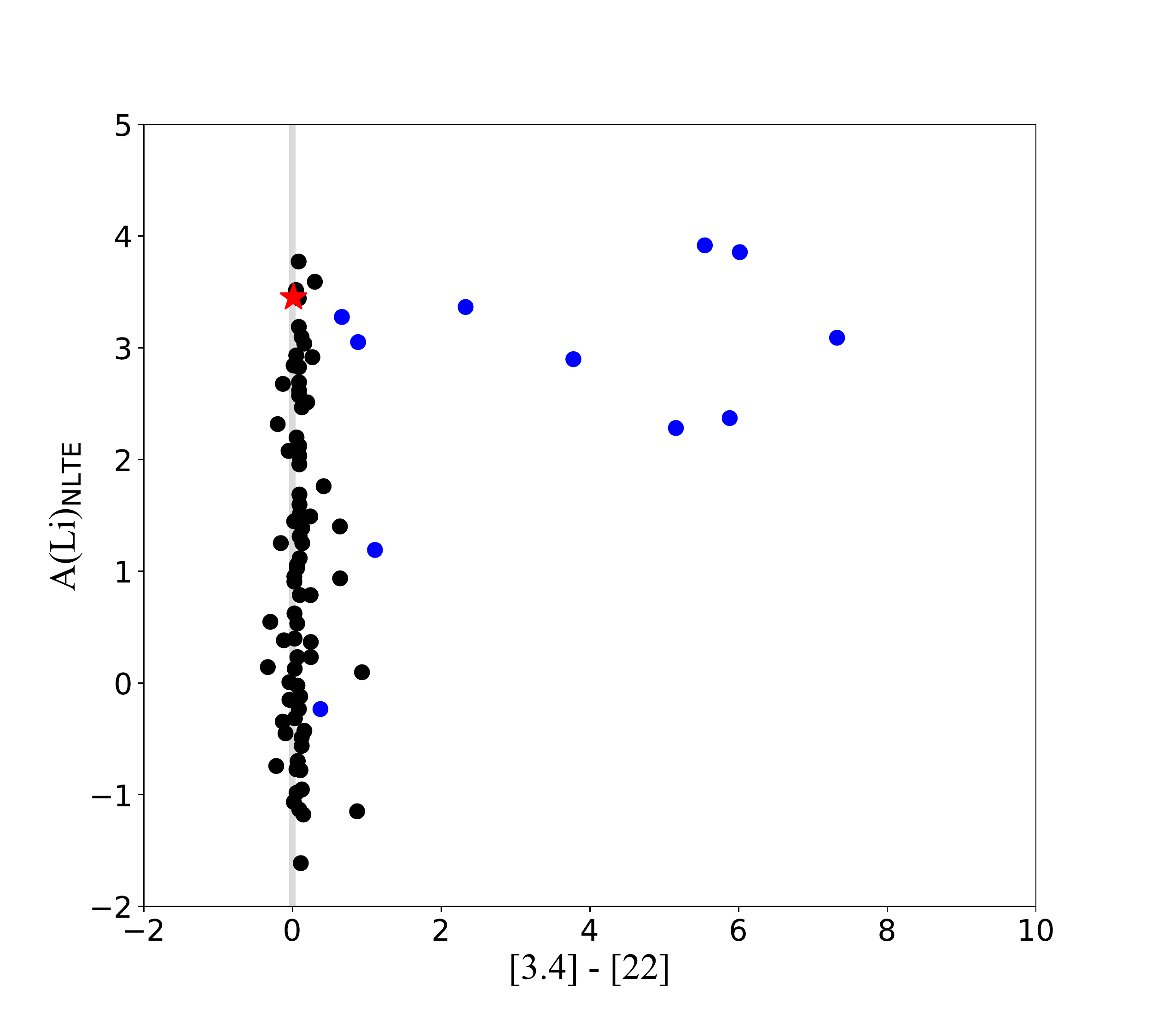}
      \caption{(Li)$_\mathrm{NLTE}$ vs. [3.4]-[22] for the sample stars of \citet{Rebull15}. The vertical line at [3.4]-[22]=0 indicates the photospheric locus. The blue points denote sources identified with IR excess by \citet{Rebull15}, while the black ones are for these with no obvious IR excess. The red star is our target.}
         \label{figure4}
 \end{figure}

It was suggested by \citet{dela97}  that enhanced lithium is associated with IR excesses for K giants. In their model, the normal Li-poor K giants would turn out to be Li-rich caused by the internal mixing during a short period ($\sim10^5$yr), which would result in an expansion of the circumstellar shell along with the mass loss \citep[see also][]{Drake02,dela15}. To check the scenario for our target, we extracted IR magnitudes of TYC\,3251-581-1 at [3.4, 4.6, 12 and 22 $\mu m$ bands] from WISE \citep{Cutri13}, following the procedure described by \citet{Rebull15}, and plot A(Li)$_\mathrm{NLTE}$ as a function of [3.4]-[22] in Fig. \ref{figure4}. It shows our target with the 176 red giants from the cleanest possible sample of \citet{Rebull15}. Here [3.4]-[22] is defined as the difference in magnitudes of these two bands. \citet{Rebull15} also defined $\chi_\mathrm{[3.4],[22]}$ as:\\
\newline
    $\chi_\mathrm{[3.4],[22]}=\frac{([3.4]-[22])_\mathrm{observed}-([3.4]-[22])_\mathrm{predicted}}{\sigma_\mathrm{([3.4]-[22])}}$\\
where ${\sigma_\mathrm{([3.4]-[22])}}$ is quadratic sum of the errors on 3.4 $\mu m$ and 22 $\mu m$. $([3.4]-[22])_\mathrm{predicted}$ is expected to be 0.0 for the K giant, which implies that the giant is without any circumstellar dust. $\chi_\mathrm{[3.4]-[22]} \ge 3.0 $ is set as the criterion for significant IR excess. The $\chi_\mathrm{[3.4],[22]}$ for TYC\,3251-581-1 is 0.36, much lower than the value of criterion, which clearly shows that TYC\,3251-581-1 does not have IR excess at [3.4]-[22]. The result is consistent with that of \citet{Rebull15}, who shows in their Fig. 19 that the slow rotators are not likely to have IR excess. Furthermore, if Li-rich giants experienced mass-loss in the form of gas, their $\mathrm{H}\alpha$ line profiles are likely asymmetrical or shifted \citep{Reddy02, Meszaros09}. There is no evidence for either asymmetry or a shift in the spectra of our target. In this case, the connection between infrared excess and the Li enhancement seems to be disfavoured by our data.

\begin{figure}
   \centering
   \includegraphics[width=9.0cm]{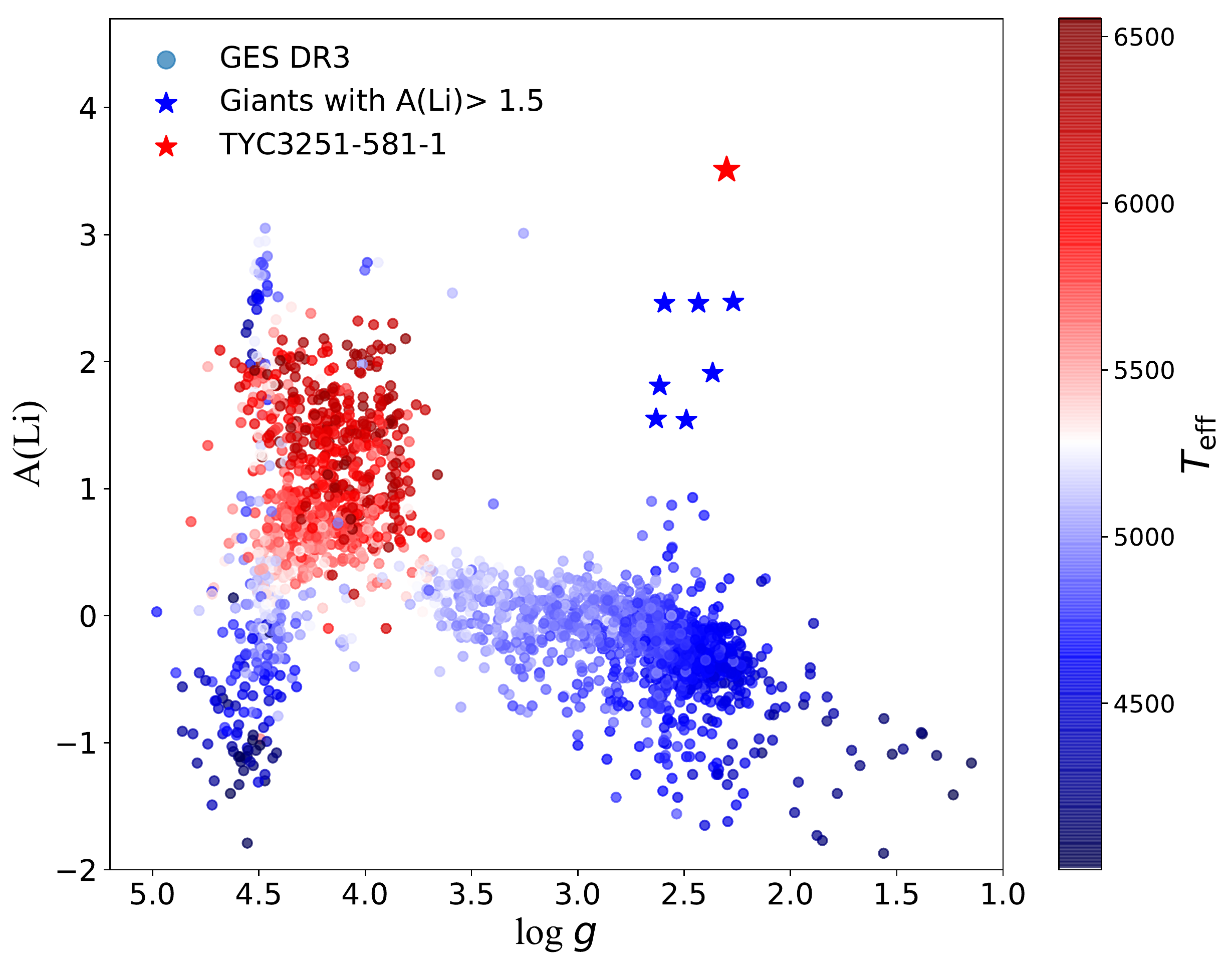}
      \caption{A(Li) evolution plotted by $\log g$ in the Gaia-ESO DR3 with [Fe/H]= $-0.2 \sim 0.0$. The magenta star shows the TYC 3251-581-1, while the blue stars are the Li-rich giants discovered by \citet{Casey16}.}
   \label{figure5}
 \end{figure}

\begin{figure*}
    \centering
    \includegraphics[width=17cm]{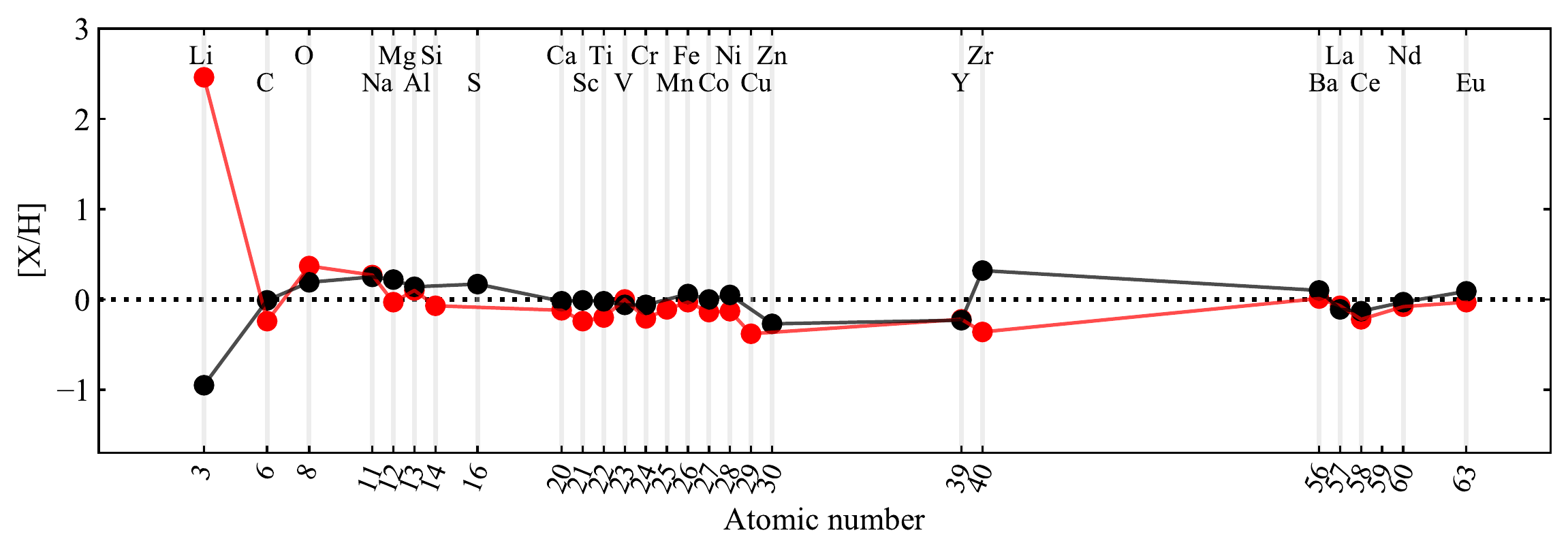}
      \caption{Comparison of detailed chemical abundances between the super Li-rich giant TYC 3251-581-1 and the Li normal giant 17562024-4134502, which are shown in red and black respectively.}
         \label{figure6}
 \end{figure*}

To intuitively inspect the Li abundance varying with the evolutionary stage, Fig. \ref{figure5} (similar as the Fig. 3 of \citealp{Aguilera16b}) shows A(Li) versus $\log g$ for stars in the Gaia-ESO DR3\footnote[3]{https://www.gaia-eso.eu} with the metallicity ranged from -0.2 to 0.0. There are seven giants with Li abundance higher than the conventional criterion (A(Li) $\ge$ 1.5 dex) of Li-rich. Three of them with A(Li) $> 2.0$ locate at the evolutionary stage before the RGB bump, they can be explained by the external scenario \citep{Casey16}. Given similar metallicity, TYC\,3251-581-1 pops up around the RGB bump with a much higher Li abundance than the maximum value of 2.2 dex predicted by \citet{Aguilera16a}. In order to compare with the Li normal giant at the chemical view, we selected the star 17562024-4134502 from the Gaia-ESO DR3 which has similar atmospheric parameters to TYC 3251-581-1. A comparison of element-to-element abundances between the two stars is illustrated in Fig. \ref{figure6}. The detailed chemical abundances of star 17562024-4134502 from Li to Eu are presented in the Gaia-ESO DR3 catalogue. If the mass transfers from the evolved AGB companion to the star, the s-process elements (e.g. Ba) are expected to be enhanced together with the Li. The Ba abundance is approximately solar, and agrees with the prediction by \citet{Bisterzo17} for the Galactic thin disc star. Thus, the external scenario of mass transfer seems to be inappropriate for our target. Except for Li, no evident differences on element abundances are found. It is consistent with the conclusion of other works \citep{Martell13, Casey16}, that the abnormal high Li abundance is unlikely connected with other elemental abundances.

A large Li-rich sample was displayed in a plot of A(Li) versus $^{12}\rm{C}/^{13}\rm{C}$ by \citealp{Rebull15} (their Figure 21) and only three Li-rich giants are located at the region that has A(Li) exceeding the meteoritic value and $^{12}\rm{C}/^{13}\rm{C} \le 15$. In the metallicity range of our target, only one star from \citeauthor{Rebull15} (HD\,77361, also shown in Fig. \ref{figure3}) is also in the RGB bump region of the Hertzsprung-Russell diagram. TYC\,3251-581-1 resembles HD\,77361 in that it has high Li abundance and also low $^{12}\rm{C}/^{13}\rm{C}$ ratio, as well as similar atmospheric parameters \citep{Lyubimkov15}. It does imply that the similar Li enrichment mechanisms might affect both of them. HD\,77361 exhibits no IR excess \citep{Rebull15} with the low rotational velocity of 4.5 $\rm{km\,s^{-1}}$ \citep{Lyubimkov15}, it seems to be in accordance with the suggestion that HD\,77361 is likely to internally produce substantial Li \citep{Kumar09}. HD\,77361 also belongs to the thin disc population from the spatial velocities $(\rm{U}_\mathrm{LSR}, \rm{V}_\mathrm{LSR}, \rm{W}_\mathrm{LSR})$ = (46, 24, -5) $\rm{km\,s^{-1}}$. From the point of \citet{Guiglion16}, the thin disc stars could be suffered higher Li enrichment than the thick disc stars due to the contribution of low mass stars. However, the contamination from the external sources is not expected to happen with respect to the low rotational velocity and non-IR excess. The decreased [C/Fe] and enhanced [N/Fe] of TYC\,3251-581-1 imply that an extra mixing should have taken place before or at the bump, and the relatively low $^{12}\rm{C}/^{13}\rm{C}$ is in agreement with this suggestion. Interestingly, the $^{12}\rm{C}/^{13}\rm{C}\sim9.0$ of TYC\,3251-581-1 is lower than that of the prediction ($19\sim23$) of the extra mixing (e.g. thermohaline mixing \citealp{Charbonnel10}; $\delta \mu \text{-} \rm{mixing}$ \citealp{Eggleton08}) for Li-rich at the end of FDU. More observations, such as asteroseismology information, are required to provide more constraints to explain this anomaly.

\section{Conclusions}
In summary of our analysis, TYC\,3251-581-1 lies close to the RGB bump evolution stage, as shown in Fig. \ref{figure3}, where extra-mixing can happen and induce Cameron-Fowler mechanism to enhance lithium. Our derived low carbon isotopic ratio, along with the low carbon abundance and enhanced nitrogen abundance suggest that the deep mixing have taken place in this star. However, for the super Li-rich giant with a low $^{12}\rm{C}/^{13}\rm{C}$ near the bump, the trigger of extra mixing is still ambiguous; further studies are needed to better understand the mechanisms of Li enhancement.

\begin{table}
\tabcolsep30pt
\scriptsize
\begin{center}
\caption[2]{Derived chemical abundances for TYC 3251-581-1 \label{tab2}}
\begin{tabular}{lc}
\hline\hline\noalign{\smallskip}
 Species &  Value $\pm \sigma$ \tablefootmark{a}\\
\noalign{\smallskip}
\hline\noalign{\smallskip}

$\rm{A(Li)}_\mathrm{LTE} $\ \; (6708 \AA)              &     $3.68 \pm0.26$\\
$\rm{A(Li)}_\mathrm{NLTE}$\    (6708 \AA)              &     $3.48 \pm0.19$\\
$\rm{A(Li)}_\mathrm{LTE} $\ \; (6104 \AA)              &     $3.38 \pm0.13$\\
$\rm{A(Li)}_\mathrm{NLTE}$\    (6104 \AA)              &     $3.53 \pm0.13$\\
$^{12}\rm{C}/^{13}\rm{C}$                              &     $9.0  \pm1.5$\\
$[\ion{C }{i}/\rm{Fe}]$                                &     $-0.34\pm0.10$\\
$[\ion{N }{i}/\rm{Fe}]$                                &     $0.33 \pm0.01$\\
$[\ion{O }{i}/\rm{Fe}]$                                &     $0.20$\\
$[\ion{Na}{i}/\rm{Fe}]$                                &     $0.25 \pm0.02$\\
$[\ion{Mg}{i}/\rm{Fe}]$                                &     $0.06 \pm0.04$\\
$[\ion{Al}{i}/\rm{Fe}]$                                &     $0.13 \pm0.02$\\
$[\ion{Si}{i}/\rm{Fe}]$                                &     $0.03 \pm0.06$\\
$[\ion{Ca}{i}/\rm{Fe}]$                                &     $-0.08\pm0.03$\\
$[\ion{Sc}{ii}/\rm{Fe}]$                               &     $0.03 \pm0.05$\\
$[\ion{Ti}{i}/\rm{Fe}]$                                &     $-0.16\pm0.07$\\
$[\ion{Ti}{ii}/\rm{Fe}]$                               &     $-0.02\pm0.07$\\
$[\ion{V }{i}/\rm{Fe}]$                                &     $0.18 \pm0.05$\\
$[\ion{Cr}{i}/\rm{Fe}]$                                &     $-0.09\pm0.03$\\
$[\ion{Mn}{i}/\rm{Fe}]$                                &     $-0.10\pm0.04$\\
$[\ion{Co}{i}/\rm{Fe}]$                                &     $0.02 \pm0.01$\\
$[\ion{Ni}{i}/\rm{Fe}]$                                &     $0.01 \pm0.05$\\
$[\ion{Cu}{i}/\rm{Fe}]$                                &     $-0.32\pm0.01$\\
$[\ion{Y }{ii}/\rm{Fe}]$                               &     $-0.10\pm0.03$\\
$[\ion{Zr}{i}/\rm{Fe}]$                                &     $-0.15\pm0.05$\\
$[\ion{Ba}{ii}/\rm{Fe}]$                               &     $0.09 \pm0.02$\\
$[\ion{La}{ii}/\rm{Fe}]$                               &     $0.08 $\\
$[\ion{Ce}{i}/\rm{Fe}]$                                &     $-0.01\pm0.01$\\
$[\ion{Nd}{ii}/\rm{Fe}]$                               &     $0.06 \pm0.01$\\
$[\ion{Eu}{ii}/\rm{Fe}]$                               &     $0.10 $\\

\noalign{\smallskip} \hline
\end{tabular}
  \tablefoot{\tablefoottext{a}{For Li, the $\sigma$ denotes the quadratic sum of uncertainties due to the atmospheric parameters, and $\sigma$ is the uncertainty of the spectral synthesis fit for $^{12}\rm{C}/^{13}\rm{C}$ and N, while it represents the standard deviation of the elemental abundances from the individual lines for the remaining elements.}}
\end{center}
\end{table}

\begin{table*}
\centering
\caption{Changes in NLTE correction for the two Li lines}
\label{table: 3}
\begin{tabular}{ccccccc}
\hline\hline
 & \citet{shi07} & \citet{Lind09} & $\Delta T_\mathrm{eff} $ (80 K) & $\Delta \log g $ (0.2 dex) & $\Delta$[Fe/H] (0.1 dex) & $\Delta \xi_\mathrm{t}$ 0.2 ($\mathrm{km\,s^{-1}}$) \\
\hline
$ \Delta_\mathrm{NLTE}$ 6708 \AA & -0.20 & -0.22 & -0.27 & -0.33 & -0.01 & -0.01 \\
$ \Delta_\mathrm{NLTE}$ 6104 \AA &  0.15 &  0.18 &  0.14 &  0.13 &  0.14 &  0.13 \\
\hline
\end{tabular}
\end{table*}

\begin{acknowledgements}
This research was supported by National Key Basic Research Programme of China 2014CB845700, and by the National Natural Science Foundation of China under grant Nos. 11603037 and 11473033. This work is supported by the Astronomical Big Data Joint Research Center, co-founded by the National Astronomical Observatories, Chinese Academy of Sciences and the Alibaba Cloud. K.P. acknowledges supports from the Mt. Cuba Astronomical Foundation Grant and from Center for Astronomical Mega$_\mathrm{science}$, CAS.

Guoshoujing Telescope (the Large Sky Area Multi-Object Fiber Spectroscopic Telescope LAMOST) is a National Major Scientific Project built by the Chinese Academy of Sciences. Funding for the project has been provided by the National Development and Reform Commission. LAMOST is operated and managed by the National Astronomical Observatories, Chinese Academy of Sciences. This publication makes use of data products from the Wide field Infrared Survey Explorer, which is a joint project of the University of California, Los Angeles, and the Jet Propulsion Laboratory/California Institute of Technology, funded by the National Aeronautics and Space Administration. Based on data products from observations made with ESO Telescopes at the La Silla Paranal Observatory under programme ID 188.B-3002.
\end{acknowledgements}

  \bibliographystyle{aa}
  \bibliography{zhou}

\begin{thebibliography}{62}
\expandafter\ifx\csname natexlab\endcsname\relax\def\natexlab#1{#1}\fi

\bibitem[{{Adam{\'o}w} {et~al.}(2015){Adam{\'o}w}, {Niedzielski}, {Villaver},
  {Wolszczan}, {Kowalik}, {Nowak}, {Adamczyk}, \&
  {Deka-Szymankiewicz}}]{Adamow15}
{Adam{\'o}w}, M., {Niedzielski}, A., {Villaver}, E., {et~al.} 2015, \aap, 581,
  A94

\bibitem[{{Adam{\'o}w} {et~al.}(2014){Adam{\'o}w}, {Niedzielski}, {Villaver},
  {Wolszczan}, \& {Nowak}}]{Adamow14}
{Adam{\'o}w}, M., {Niedzielski}, A., {Villaver}, E., {Wolszczan}, A., \&
  {Nowak}, G. 2014, \aap, 569, A55

\bibitem[{{Adibekyan} {et~al.}(2011){Adibekyan}, {Santos}, {Sousa}, \&
  {Israelian}}]{Adibekyan11}
{Adibekyan}, V.~Z., {Santos}, N.~C., {Sousa}, S.~G., \& {Israelian}, G. 2011,
  \aap, 535, L11

\bibitem[{{Adibekyan} {et~al.}(2012){Adibekyan}, {Sousa}, {Santos}, {Delgado
  Mena}, {Gonz{\'a}lez Hern{\'a}ndez}, {Israelian}, {Mayor}, \&
  {Khachatryan}}]{Adibekyan12}
{Adibekyan}, V.~Z., {Sousa}, S.~G., {Santos}, N.~C., {et~al.} 2012, \aap, 545,
  A32

\bibitem[{{Aguilera-G{\'o}mez}
  {et~al.}(2016{\natexlab{a}}){Aguilera-G{\'o}mez}, {Chanam{\'e}},
  {Pinsonneault}, \& {Carlberg}}]{Aguilera16b}
{Aguilera-G{\'o}mez}, C., {Chanam{\'e}}, J., {Pinsonneault}, M.~H., \&
  {Carlberg}, J.~K. 2016{\natexlab{a}}, \apjl, 833, L24

\bibitem[{{Aguilera-G{\'o}mez}
  {et~al.}(2016{\natexlab{b}}){Aguilera-G{\'o}mez}, {Chanam{\'e}},
  {Pinsonneault}, \& {Carlberg}}]{Aguilera16a}
{Aguilera-G{\'o}mez}, C., {Chanam{\'e}}, J., {Pinsonneault}, M.~H., \&
  {Carlberg}, J.~K. 2016{\natexlab{b}}, \apj, 829, 127

\bibitem[{{Alexander}(1967)}]{Alexander1967}
{Alexander}, J.~B. 1967, The Observatory, 87, 238

\bibitem[{{Alexeeva} \& {Mashonkina}(2015)}]{Alexeeva15}
{Alexeeva}, S.~A. \& {Mashonkina}, L.~I. 2015, \mnras, 453, 1619

\bibitem[{{Alonso} {et~al.}(1999){Alonso}, {Arribas}, \&
  {Mart{\'{\i}}nez-Roger}}]{Alonso99}
{Alonso}, A., {Arribas}, S., \& {Mart{\'{\i}}nez-Roger}, C. 1999, \aaps, 140,
  261

\bibitem[{{Asplund} {et~al.}(2009){Asplund}, {Grevesse}, {Sauval}, \&
  {Scott}}]{Asplund09}
{Asplund}, M., {Grevesse}, N., {Sauval}, A.~J., \& {Scott}, P. 2009, \araa, 47,
  481

\bibitem[{{Bisterzo} {et~al.}(2017){Bisterzo}, {Travaglio}, {Wiescher},
  {K{\"a}ppeler}, \& {Gallino}}]{Bisterzo17}
{Bisterzo}, S., {Travaglio}, C., {Wiescher}, M., {K{\"a}ppeler}, F., \&
  {Gallino}, R. 2017, \apj, 835, 97

\bibitem[{{Brown} {et~al.}(1989){Brown}, {Sneden}, {Lambert}, \&
  {Dutchover}}]{Brown89}
{Brown}, J.~A., {Sneden}, C., {Lambert}, D.~L., \& {Dutchover}, Jr., E. 1989,
  \apjs, 71, 293

\bibitem[{{Cameron} \& {Fowler}(1971)}]{Cameron71}
{Cameron}, A.~G.~W. \& {Fowler}, W.~A. 1971, \apj, 164, 111

\bibitem[{{Carlberg} {et~al.}(2012){Carlberg}, {Cunha}, {Smith}, \&
  {Majewski}}]{Carlberg12}
{Carlberg}, J.~K., {Cunha}, K., {Smith}, V.~V., \& {Majewski}, S.~R. 2012,
  \apj, 757, 109

\bibitem[{{Carlberg} {et~al.}(2010){Carlberg}, {Smith}, {Cunha}, {Majewski}, \&
  {Rood}}]{Carlberg10}
{Carlberg}, J.~K., {Smith}, V.~V., {Cunha}, K., {Majewski}, S.~R., \& {Rood},
  R.~T. 2010, \apjl, 723, L103

\bibitem[{{Casey} {et~al.}(2016){Casey}, {Ruchti}, {Masseron}, {Randich},
  {Gilmore}, {Lind}, {Kennedy}, {Koposov}, {Hourihane}, {Franciosini}, {Lewis},
  {Magrini}, {Morbidelli}, {Sacco}, {Worley}, {Feltzing}, {Jeffries},
  {Vallenari}, {Bensby}, {Bragaglia}, {Flaccomio}, {Francois}, {Korn},
  {Lanzafame}, {Pancino}, {Recio-Blanco}, {Smiljanic}, {Carraro}, {Costado},
  {Damiani}, {Donati}, {Frasca}, {Jofr{\'e}}, {Lardo}, {de Laverny}, {Monaco},
  {Prisinzano}, {Sbordone}, {Sousa}, {Tautvai{\v s}ien{\.e}}, {Zaggia},
  {Zwitter}, {Delgado Mena}, {Chorniy}, {Martell}, {Silva Aguirre}, {Miglio},
  {Chiappini}, {Montalban}, {Morel}, \& {Valentini}}]{Casey16}
{Casey}, A.~R., {Ruchti}, G., {Masseron}, T., {et~al.} 2016, \mnras, 461, 3336

\bibitem[{{Charbonnel} \& {Balachandran}(2000)}]{Charbonnel00}
{Charbonnel}, C. \& {Balachandran}, S.~C. 2000, \aap, 359, 563

\bibitem[{{Charbonnel} \& {Lagarde}(2010)}]{Charbonnel10}
{Charbonnel}, C. \& {Lagarde}, N. 2010, \aap, 522, A10

\bibitem[{{Charbonnel} \& {Zahn}(2007)}]{Charbonnel07}
{Charbonnel}, C. \& {Zahn}, J.-P. 2007, \aap, 467, L15

\bibitem[{{Cui} {et~al.}(2012){Cui}, {Zhao}, {Chu}, {Li}, {Li}, {Zhang}, {Su},
  {Yao}, {Wang}, {Xing}, {Li}, {Zhu}, {Wang}, {Gu}, {Luo}, {Xu}, {Zhang},
  {Liu}, {Zhang}, {Yang}, {Cao}, {Chen}, {Chen}, {Chen}, {Chen}, {Chu}, {Feng},
  {Gong}, {Hou}, {Hu}, {Hu}, {Hu}, {Jia}, {Jiang}, {Jiang}, {Jiang}, {Jin},
  {Li}, {Li}, {Li}, {Liu}, {Liu}, {Lu}, {Mao}, {Men}, {Qi}, {Qi}, {Shi},
  {Tang}, {Tao}, {Wang}, {Wang}, {Wang}, {Wang}, {Wang}, {Wang}, {Wang},
  {Wang}, {Wang}, {Wang}, {Wang}, {Wang}, {Xu}, {Xu}, {Yang}, {Yu}, {Yuan},
  {Yuan}, {Zhai}, {Zhang}, {Zhang}, {Zhang}, {Zhao}, {Zhou}, {Zhou}, {Zhu}, \&
  {Zou}}]{Cui12}
{Cui}, X.-Q., {Zhao}, Y.-H., {Chu}, Y.-Q., {et~al.} 2012, Research in Astronomy
  and Astrophysics, 12, 1197

\bibitem[{{Cutri} \& {et al.}(2013)}]{Cutri13}
{Cutri}, R.~M. \& {et al.} 2013, VizieR Online Data Catalog, 2328

\bibitem[{{de la Reza} {et~al.}(1997){de la Reza}, {Drake}, {da Silva},
  {Torres}, \& {Martin}}]{dela97}
{de la Reza}, R., {Drake}, N.~A., {da Silva}, L., {Torres}, C.~A.~O., \&
  {Martin}, E.~L. 1997, \apjl, 482, L77

\bibitem[{{de la Reza} {et~al.}(2015){de la Reza}, {Drake}, {Oliveira}, \&
  {Rengaswamy}}]{dela15}
{de la Reza}, R., {Drake}, N.~A., {Oliveira}, I., \& {Rengaswamy}, S. 2015,
  \apj, 806, 86

\bibitem[{{Drake} {et~al.}(2002){Drake}, {de la Reza}, {da Silva}, \&
  {Lambert}}]{Drake02}
{Drake}, N.~A., {de la Reza}, R., {da Silva}, L., \& {Lambert}, D.~L. 2002,
  \aj, 123, 2703

\bibitem[{{Eggleton} {et~al.}(2008){Eggleton}, {Dearborn}, \&
  {Lattanzio}}]{Eggleton08}
{Eggleton}, P.~P., {Dearborn}, D.~S.~P., \& {Lattanzio}, J.~C. 2008, \apj, 677,
  581

\bibitem[{{Gaia Collaboration} {et~al.}(2016){Gaia Collaboration}, {Brown},
  {Vallenari}, {Prusti}, {de Bruijne}, {Mignard}, {Drimmel}, {Babusiaux},
  {Bailer-Jones}, {Bastian}, \& et~al.}]{Gaia16}
{Gaia Collaboration}, {Brown}, A.~G.~A., {Vallenari}, A., {et~al.} 2016, \aap,
  595, A2

\bibitem[{{Grupp} {et~al.}(2009){Grupp}, {Kurucz}, \& {Tan}}]{gru09}
{Grupp}, F., {Kurucz}, R.~L., \& {Tan}, K. 2009, \aap, 503, 177

\bibitem[{{Guiglion} {et~al.}(2016){Guiglion}, {de Laverny}, {Recio-Blanco},
  {Worley}, {De Pascale}, {Masseron}, {Prantzos}, \& {Mikolaitis}}]{Guiglion16}
{Guiglion}, G., {de Laverny}, P., {Recio-Blanco}, A., {et~al.} 2016, \aap, 595,
  A18

\bibitem[{{H{\o}g} {et~al.}(2000){H{\o}g}, {Fabricius}, {Makarov}, {Urban},
  {Corbin}, {Wycoff}, {Bastian}, {Schwekendiek}, \& {Wicenec}}]{Hog00}
{H{\o}g}, E., {Fabricius}, C., {Makarov}, V.~V., {et~al.} 2000, \aap, 355, L27

\bibitem[{{Iben}(1967{\natexlab{a}})}]{Iben67b}
{Iben}, Jr., I. 1967{\natexlab{a}}, \apj, 147, 650

\bibitem[{{Iben}(1967{\natexlab{b}})}]{Iben67a}
{Iben}, Jr., I. 1967{\natexlab{b}}, \apj, 147, 624

\bibitem[{{Israelian} {et~al.}(2001){Israelian}, {Santos}, {Mayor}, \&
  {Rebolo}}]{Israelian01}
{Israelian}, G., {Santos}, N.~C., {Mayor}, M., \& {Rebolo}, R. 2001, \nat, 411,
  163

\bibitem[{{Jacobson} \& {Friel}(2013)}]{Jacobson13}
{Jacobson}, H.~R. \& {Friel}, E.~D. 2013, \aj, 145, 107

\bibitem[{{Kirby} {et~al.}(2016){Kirby}, {Guhathakurta}, {Zhang}, {Hong},
  {Guo}, {Guo}, {Cohen}, \& {Cunha}}]{Kirby16}
{Kirby}, E.~N., {Guhathakurta}, P., {Zhang}, A.~J., {et~al.} 2016, \apj, 819,
  135

\bibitem[{{Kumar} \& {Reddy}(2009)}]{Kumar09}
{Kumar}, Y.~B. \& {Reddy}, B.~E. 2009, \apjl, 703, L46

\bibitem[{{Kumar} {et~al.}(2011){Kumar}, {Reddy}, \& {Lambert}}]{Kumar11}
{Kumar}, Y.~B., {Reddy}, B.~E., \& {Lambert}, D.~L. 2011, \apjl, 730, L12

\bibitem[{{Kupka} {et~al.}(1999){Kupka}, {Piskunov}, {Ryabchikova}, {Stempels},
  \& {Weiss}}]{Kupka99}
{Kupka}, F., {Piskunov}, N., {Ryabchikova}, T.~A., {Stempels}, H.~C., \&
  {Weiss}, W.~W. 1999, \aaps, 138, 119

\bibitem[{{Lebzelter} {et~al.}(2012){Lebzelter}, {Uttenthaler}, {Busso},
  {Schultheis}, \& {Aringer}}]{Lebzelter12}
{Lebzelter}, T., {Uttenthaler}, S., {Busso}, M., {Schultheis}, M., \&
  {Aringer}, B. 2012, \aap, 538, A36

\bibitem[{{Li} {et~al.}(2018){Li}, {Aoki}, {Matsuno}, {Bharat Kumar}, {Shi},
  {Suda}, \& {Zhao}}]{Li18}
{Li}, H., {Aoki}, W., {Matsuno}, T., {et~al.} 2018, \apjl, 852, L31

\bibitem[{{Li} {et~al.}(2016){Li}, {Aoki}, {Zhao}, {Honda}, {Christlieb}, \&
  {Suda}}]{Li16}
{Li}, H., {Aoki}, W., {Zhao}, G., {et~al.} 2016, in IAU Symposium, Vol. 317,
  The General Assembly of Galaxy Halos: Structure, Origin and Evolution, ed.
  A.~{Bragaglia}, M.~{Arnaboldi}, M.~{Rejkuba}, \& D.~{Romano}, 51--56

\bibitem[{{Lind} {et~al.}(2009){Lind}, {Asplund}, \& {Barklem}}]{Lind09}
{Lind}, K., {Asplund}, M., \& {Barklem}, P.~S. 2009, \aap, 503, 541

\bibitem[{{Lyubimkov} {et~al.}(2015){Lyubimkov}, {Kaminsky}, {Metlov},
  {Pavlenko}, {Poklad}, \& {Rachkovskaya}}]{Lyubimkov15}
{Lyubimkov}, L.~S., {Kaminsky}, B.~M., {Metlov}, V.~G., {et~al.} 2015,
  Astronomy Letters, 41, 809

\bibitem[{{Martell} \& {Shetrone}(2013)}]{Martell13}
{Martell}, S.~L. \& {Shetrone}, M.~D. 2013, \mnras, 430, 611

\bibitem[{{Martin} {et~al.}(1994){Martin}, {Rebolo}, {Casares}, \&
  {Charles}}]{Martin94}
{Martin}, E.~L., {Rebolo}, R., {Casares}, J., \& {Charles}, P.~A. 1994, \apj,
  435, 791

\bibitem[{{M{\'e}sz{\'a}ros} {et~al.}(2009){M{\'e}sz{\'a}ros}, {Dupree}, \&
  {Szalai}}]{Meszaros09}
{M{\'e}sz{\'a}ros}, S., {Dupree}, A.~K., \& {Szalai}, T. 2009, \aj, 137, 4282

\bibitem[{{Monaco} {et~al.}(2014){Monaco}, {Boffin}, {Bonifacio}, {Villanova},
  {Carraro}, {Caffau}, {Steffen}, {Ahumada}, {Beletsky}, \&
  {Beccari}}]{Monaco14}
{Monaco}, L., {Boffin}, H.~M.~J., {Bonifacio}, P., {et~al.} 2014, \aap, 564, L6

\bibitem[{{Monaco} {et~al.}(2011){Monaco}, {Villanova}, {Moni Bidin},
  {Carraro}, {Geisler}, {Bonifacio}, {Gonzalez}, {Zoccali}, \&
  {Jilkova}}]{Monaco11}
{Monaco}, L., {Villanova}, S., {Moni Bidin}, C., {et~al.} 2011, \aap, 529, A90

\bibitem[{{Neves} {et~al.}(2009){Neves}, {Santos}, {Sousa}, {Correia}, \&
  {Israelian}}]{Neves09}
{Neves}, V., {Santos}, N.~C., {Sousa}, S.~G., {Correia}, A.~C.~M., \&
  {Israelian}, G. 2009, \aap, 497, 563

\bibitem[{{Palmerini} {et~al.}(2011){Palmerini}, {Cristallo}, {Busso}, {Abia},
  {Uttenthaler}, {Gialanella}, \& {Maiorca}}]{Palmerini11}
{Palmerini}, S., {Cristallo}, S., {Busso}, M., {et~al.} 2011, \apj, 741, 26

\bibitem[{{Pfeiffer} {et~al.}(1998){Pfeiffer}, {Frank}, {Baumueller},
  {Fuhrmann}, \& {Gehren}}]{pfe98}
{Pfeiffer}, M.~J., {Frank}, C., {Baumueller}, D., {Fuhrmann}, K., \& {Gehren},
  T. 1998, \aaps, 130, 381

\bibitem[{{Pietrinferni} {et~al.}(2004){Pietrinferni}, {Cassisi}, {Salaris}, \&
  {Castelli}}]{Pietrinferni04}
{Pietrinferni}, A., {Cassisi}, S., {Salaris}, M., \& {Castelli}, F. 2004, \apj,
  612, 168

\bibitem[{{Rebull} {et~al.}(2015){Rebull}, {Carlberg}, {Gibbs}, {Deeb},
  {Larsen}, {Black}, {Altepeter}, {Bucksbee}, {Cashen}, {Clarke}, {Datta},
  {Hodgson}, \& {Lince}}]{Rebull15}
{Rebull}, L.~M., {Carlberg}, J.~K., {Gibbs}, J.~C., {et~al.} 2015, \aj, 150,
  123

\bibitem[{{Reddy} {et~al.}(2006){Reddy}, {Lambert}, \& {Allende
  Prieto}}]{Reddy06}
{Reddy}, B.~E., {Lambert}, D.~L., \& {Allende Prieto}, C. 2006, \mnras, 367,
  1329

\bibitem[{{Reddy} {et~al.}(2002){Reddy}, {Lambert}, {Hrivnak}, \&
  {Bakker}}]{Reddy02}
{Reddy}, B.~E., {Lambert}, D.~L., {Hrivnak}, B.~J., \& {Bakker}, E.~J. 2002,
  \aj, 123, 1993

\bibitem[{{Reetz}(1991)}]{Reetz91}
{Reetz}, J.~K. 1991, Diploma Thesis, Universit\"{a}t M\"{u}nchen

\bibitem[{{Sackmann} \& {Boothroyd}(1999)}]{Sackmann99}
{Sackmann}, I.-J. \& {Boothroyd}, A.~I. 1999, \apj, 510, 217

\bibitem[{{Shi} {et~al.}(2007){Shi}, {Gehren}, {Zhang}, {Zeng}, \&
  {Zhao}}]{shi07}
{Shi}, J.~R., {Gehren}, T., {Zhang}, H.~W., {Zeng}, J.~L., \& {Zhao}, G. 2007,
  \aap, 465, 587

\bibitem[{{Siess} \& {Livio}(1999)}]{Siess99}
{Siess}, L. \& {Livio}, M. 1999, \mnras, 308, 1133

\bibitem[{{Silva Aguirre} {et~al.}(2014){Silva Aguirre}, {Ruchti}, {Hekker},
  {Cassisi}, {Christensen-Dalsgaard}, {Datta}, {Jendreieck}, {Jessen-Hansen},
  {Mazumdar}, {Mosser}, {Stello}, {Beck}, \& {de Ridder}}]{Silva14}
{Silva Aguirre}, V., {Ruchti}, G.~R., {Hekker}, S., {et~al.} 2014, \apjl, 784,
  L16

\bibitem[{{Wallerstein} \& {Sneden}(1982)}]{Wallerstein82}
{Wallerstein}, G. \& {Sneden}, C. 1982, \apj, 255, 577

\bibitem[{{Zhao} {et~al.}(2016){Zhao}, {Mashonkina}, {Yan}, {Alexeeva},
  {Kobayashi}, {Pakhomov}, {Shi}, {Sitnova}, {Tan}, {Zhang}, {Zhang}, {Zhou},
  {Bolte}, {Chen}, {Li}, {Liu}, \& {Zhai}}]{Zhao16}
{Zhao}, G., {Mashonkina}, L., {Yan}, H.~L., {et~al.} 2016, \apj, 833, 225

\bibitem[{{Zhao} {et~al.}(2012){Zhao}, {Zhao}, {Chu}, {Jing}, \&
  {Deng}}]{Zhao12}
{Zhao}, G., {Zhao}, Y.-H., {Chu}, Y.-Q., {Jing}, Y.-P., \& {Deng}, L.-C. 2012,
  Research in Astronomy and Astrophysics, 12, 723

\end{thebibliography}
\end{document}